# Single microparticles mass measurement using an AFM cantilever resonator

**Abstract**: *In this work is presented a microbalance for single microparticle sensing based on resonating AFM cantilever. The variation of the resonator eigenfrequency is related to the particle mass positioned at the free apex of the cantilever. An all-digital phase locked loop (PLL) control system is developed to detect the variations in cantilever eigenfrequency. Two particle populations of different materials are used in the experimental test, demonstrating a mass sensitivity of 15 Hz/pg in ambient conditions. Thereby it is validated the possibility of developing an inexpensive, portable and sensitive microbalance for point-mass sensing.*

Marco Mauro,[a]* Raffaele Battaglia,[a] Gianluca Ferrini,[a] Roberto Puglisi,[b] Donatella Balduzzi[b] and Andrea Galli[b]

[a]Novaetech S.r.l., Via J. F. Kennedy n.5, 80125, Napoli, Italy. E-mail: mauro@novaetech.it; Fax: +39 081 011180; Tel: +39 081 056850

[b]Istituto Sperimentale Italiano "Lazzaro Spallanzani", Loc. La Quercia – Rivolta d'Adda, 26027, Cremona, Italy. Fax: +39 0363 37047981; Tel: +39 0363 78883

The development of sensors for mass measurement with high sensitivity is of great interest in a wide range of applications. Nowadays it is well-demonstrated that micromechanical resonators, in the typical form of cantilever or bridge, are very effective tools in this field [1 – 2], since the first pioneering application was presented by using a standard AFM cantilever [3]. The most recent system shows the detection of masses in the range of zeptogram [4]. Typically, the sensitivity of the mass detection is enhanced by scaling-down the dimensions of the resonator to nano-regime, or by using higher modes of vibration. Furthermore, the recent application of micromechanical resonators for detection of air-borne micro - nanoparticle shows great promises in this field [5 – 6]. In this paper it is investigated the response of a commercially available AFM cantilever caused by the deposition of single microparticles, using an all-digital phase locked loop control system. The main goal of this work is to demonstrate the possibility of developing an inexpensive microbalance for single point-mass sensing based on standard AFM cantilever.

AFM cantilever is a micromechanical device consisting of a single clamped thin beam with the other end free to vibrate. The eigenfrequency $f_0$ of the first bending mode of a resonating cantilever in vacuum is given by [1]:

$$f_0 = \frac{\lambda^2}{2\pi} \cdot \frac{t}{l} \sqrt{\frac{E}{12\rho}}, \qquad (1)$$

where $\lambda = 1.8751$ is the fundamental modal constant, $\rho$ is the mass density, $E$ is the Young's modulus, $t$ is the thickness and $l$ is the length of the cantilever. It is commonly used to write the equation (1) in the form of a one-dimensional harmonic oscillator:

$$f_0 = \frac{1}{2\pi} \sqrt{\frac{k_{eff}}{m_{eff}}}, \qquad (2)$$

where $k_{eff}$ and $m_{eff}$ denote effective spring constant and effective mass. For the first bending mode of a cantilever the effective spring constant and the effective mass are given by:

$$m_{eff} = 0.24 \cdot m_0, \quad m_0 = \rho l w t, \quad k_{eff} = \frac{E t^3 w}{4l} \qquad (3)$$

# Single microparticles mass measurement using an AFM cantilever resonator

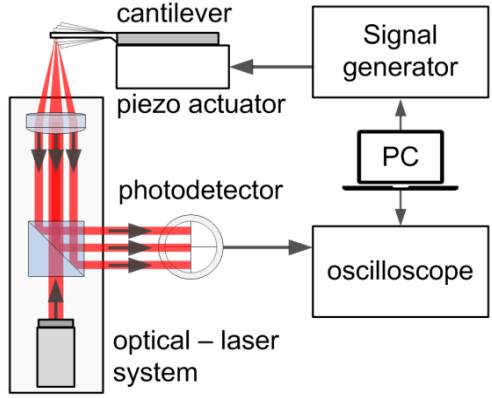

**Figure 1** Schematic of the standard AFM optical readout setup for mass measurement. The cantilever is mechanically clamped to a piezoelectric actuator driven by a signal generator. A collimated laser beam is focused using an aspheric lens on the free vibrating apex of the cantilever with a spot diameter of about 10 μm, inferior to the width of the cantilever to avoid diffraction noise. The focused beam is reflected by the aluminium-coated surface of the cantilever. The reflected beam is finally revealed by a 4-quadrant photodetector. An oscilloscope acquires the voltage signals which are fed to the PC-software for signal processing.

The deposition of a small mass $\Delta m$ compared to that of the cantilever causes the change in the eigenfrequency $\Delta f$. If the stiffness remains unchanged, the eigenfrequency variation is given by the first order approximation equation:

$$\Delta f = -2\pi^2 \cdot \frac{f_0^3}{k_{eff}} \Delta m \tag{4}$$

The equation (4) directly relates the variation in cantilever eigenfrequency to an added mass positioned at the free apex. The sensitivity $S$ of the cantilever mass sensor is defined as the ratio between the variation in eigenfrequency and the added mass. Using the equation (4) it is straightforward to calculate the sensitivity of the sensor if it is known the dimensions and the material of the cantilever.

In Fig.1 it is shown the standard AFM optical readout setup, which is used for mass measurement [7]. The mechanical motion of the cantilever is transduced in an analog electrical signal by means of a laser beam and a photodetector. The entire experimental setup is handled by custom software designed in LabVIEW development environment. The vibration of the cantilever is induced by the piezoelectric actuator PL055, *Physik Instrumente*, which is mechanically clamped to the cantilever chip. The piezoelectric actuator is driven by the signal generator, which is handled by the LabVIEW software. The reflected beam is detected by using the quadrant photodetector QD50-0-SD *OSI Optoelectronics*, which converts the photoelectric current in an oscillating voltage signal. The analog signals are acquired by using a 2-channels oscilloscope, with a time base $T = 25\ \mu s$ and a record length of 2500 samples. The LabVIEW software fetches the waveforms and processes the time signals. The software performs a Fast Fourier Transform (FFT) of the time signals to detect the frequency and phase lag.

An all-digital phase locked loop (PLL) control system is developed in LabVIEW for the measurement of the variations in eigenfrequency [8]. The PLL is designed by considering a driven damped one-dimensional harmonic oscillator model for the first cantilever bending mode. The cantilever eigenfrequency can be detected by controlling the phase lag between the driving vibration and the cantilever steady state vibration. The phase lag $\phi$ is given by [9]:

$$\tan\phi = \frac{\kappa}{\pi} \cdot \frac{f}{f^2 - f_0^2} \tag{5}$$

where $\kappa$ is the is the damping coefficient and $f$ is the driving frequency. By locking the phase lag to $\phi_0 = -\pi/2$ it is possible to detect the eigenfrequency of the cantilever whatever is the value of the damping coefficient. Moreover the PLL control system is able to track in real-time the variations in eigenfrequency caused by an added mass. The driving vibration $u_i(t)$ and resonator vibration $u_o(t)$ are defined as:

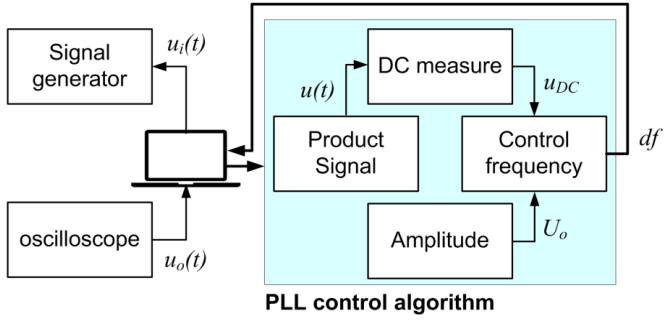

**Figure 2** Schematic diagram of the all-digital PLL control system for detecting and tracking the cantilever eigenfrequency. The PLL processes the driving $u_i(t)$ and the cantilever vibrations $u_o(t)$ to calculate the frequency control value $df$. The feedback information is fed to the signal generator in order to change the driving frequency until the set point of the phase lag is reached.

$$u_i(t) = U_i \cos(2\pi f \cdot t), \quad u_o(t) = U_o \cos(2\pi f \cdot t + \phi) \tag{6}$$

The phase lag is chosen in that way because the resonator vibration is always lagging behind the driving vibration. The product signal $u(t) = u_i(t) \cdot u_o(t)$ is the sum of a continuous $u_{DC}$ and an alternate component $u_{AC}(t)$ :

$$u_{DC} = \frac{1}{2} U_i U_o \cos \phi, \quad u_{AC} = \frac{1}{2} U_i U_o \cos(4\pi f \cdot t + \phi) \tag{7}$$

The continuous component $u_{DC}$ is proportional to $\cos \phi$ and it is null at the cantilever eigenfrequency, is positive when $f < f_0$ and negative if $f > f_0$. In order to lock the phase lag to the value $\phi_0$ the driving frequency must be changed by a value $df$ which is given by

$$df = -c_0 \left( \frac{u_{DC}}{U_o^2} \right) \tag{8}$$

where $c_0$ is the control constant parameter which depends on the particular physical system. Using equations (7), it can be verified that $df$ is independent from the damping coefficient. The schematic diagram of the PLL control system is described in Fig. 2. The oscilloscope acquires the driving and cantilever analog signals which are fed to the PC. The LabVIEW software processes the time signals acquired by the oscilloscope and it elaborates the frequency control value according to equation (8). Finally, the software handles the signal generator to change the driving frequency until the set point of the phase lag is reached. At the steady state, the PLL control system lock the phase lag to $\phi_0$ with a mean error of about 0.01 rad which roughly corresponds to a frequency resolution of 2 Hz in the experimental conditions.

One of the main goals of this work is to demonstrate the possibility of developing a point-mass sensor by using a low-cost and commercially available cantilever. To this aim, it has been selected the AFM cantilever, model ACT-TL tipless manufactured by *Applied Nanostructures*. According to the technical specifications (Table 1) and using equation (3), it is possible to calculate the theoretical sensitivity of the cantilever mass sensor: $S_{th}$ = 15 Hz/pg. In this calculation it is considered negligible the contribution coming from the aluminium nano-metric coating. Single microparticles are positioned in a selected area at the free apex

**Table 1** Geometric and physical characteristics of the cantilever sensor

| Technical data | Symbol | Tipical Value | Range |
|---|---|---|---|
| Effective spring constant | $k_{eff}$ | 40 N/m | 25 – 55 N/m |
| Resonance Frequency. | $f_0$ | 300 kHz | 200 – 400 kHz |
| Length | $l$ | 125 μm | 115 – 135 μm |
| Mean width | $w$ | 35 μm | 30 – 40 μm |
| Thickness | $t$ | 4.5 μm | 4.0 – 5.0 μm |
| Shape /Cross section | | Rectangular / Trapezoidal | |
| Material | | Single Crystal Silicon, Highly Doped | |
| Coating | | Al overall (~ 30 nm thick) | |

Single microparticles mass measurement using an AFM cantilever resonator of the cantilever by using a micro tip mounted on a manual XYZ micromanipulator (Fig. 3). After the positioning, it has been recorded the position of the microparticle and the variation in eigenfrequency caused by its mass. Finally, the microparticle is removed from the cantilever. Populations of Silica and Melamine resin microspheres, manufactured by *microparticles GmbH*, are used in the experimental tests. They are typically used for standard calibration of AFM microscope, for this reason they have a well-defined diameter and density (Table 2). According to the technical specifications, the mass of microparticles is $m_{MR}$ = 89 ± 7 pg for Melamine resin and $m_S$ = *108 ± 14* pg for Silica.

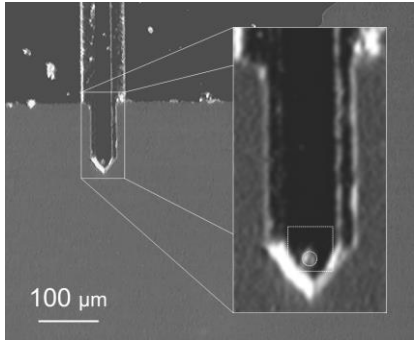

**Figure 3** Microscope image showing the AFM cantilever and the microparticle positioned at the free apex. The 3x magnified image shows the particle marked with a ring and the selected sensitive area which corresponds to a square of 20 μm length

**Table 2** Technical specifications of calibration microspheres

| Material | diameter (μm) | density (g/cm$^3$) |
|---|---|---|
| Silica | 4.78 ± 0.19 | 1.9 ± 0.1 |
| Melamine resin | 4.83 ± 0.12 | 1.51 ± 0.01 |

The distribution of eigenfrequency variations corresponding to the different populations of microparticles is shown in Fig. 4 and the data analysis are summarized in Table 3. Although the variation in eigenfrequency depends also on the particle position [1], the data processing shows that the variation in cantilever sensitivity as a function of microparticles position can be considered negligible within experimental error.

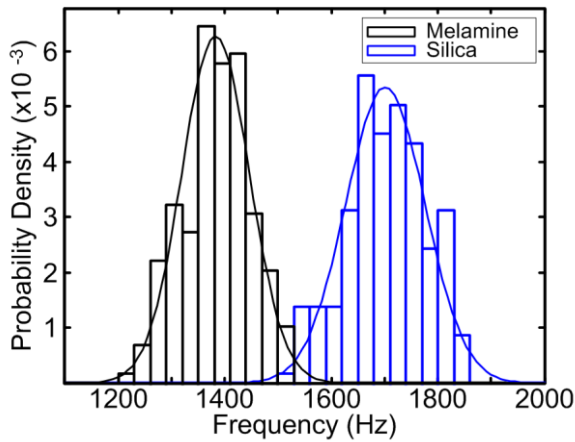

**Figure 4** Gaussian distributions fit related to eigenfrequency variations for populations of Melamine Resin and Silica microparticles. The number of observations is $N_{MR}$ = 196 and $N_S$ = 192 respectively.

**Table 3** Data analysis of two microparticle populations

| Material | Mass (pg) | Frequency (Hz) | Sensitivity (Hz/pg) |
|---|---|---|---|
| Silica | 89 ± 7 | 1383 ± 4 | 15 ± 1 |
| Melamine resin | 108 ± 14 | 1701 ± 5 | 15 ± 2 |

The experimental results demonstrate that the AFM resonating cantilever is able to measure an added mass of a single microparticle with a sensitivity of $S_{th}$ = 15 ± 2 Hz/pg in ambient conditions, according to the theoretical prediction. This work demonstrates the possibility of developing a low-cost and sensitive microbalance for single particle mass measurement using a commercially available AFM cantilever. The PLL control system designed for eigenfrequency tracking allows for real-time mass sensing. For the development of a portable smart sensor, the future work will concern the development of piezoresistitve readout scheme, which is suitable for miniaturization. Moreover, the design of a custom electronics for ADC converting and digital signal processing could be a great advantage. The development of inexpensive – smart mass sensor is of great interest in a wide range of practical applications. For example, by coating the sensitive area at the free apex of the cantilever it is possible to monitor specific chemical compounds in real-time. Furthermore, the point-mass sensing capabilities demonstrate an intriguing potential in the mass measurement of airborne microparticles.

ACKNOWLEDGEMENTS
This work was supported by the project SESSIBOV no. 9106/303/2009 MiPAAF.